\documentclass[preprint,showpacs,nopreprintnumbers,amsmath,amssymb,floatfix]{revtex4}

\usepackage{graphicx}
\usepackage{dcolumn}
\usepackage{bm}

\newcommand{\nuc}[2]{$ ^{#1}\mbox{#2}$}
\newcommand{\pbar}{$\overline{\mbox{p}}$}

\begin{document}

\preprint{APS/123-QED}

\title{Nucleon density in the nuclear periphery determined with
           antiprotonic x-rays: cadmium and tin isotopes}
\author{ 
R.~Schmidt$^1$, 
A.~Trzci{\'n}ska$^2$,
T.~Czosnyka$^2$,
T.~von~Egidy$^1$,
K.~Gulda$^3$,
F.~J.~Hartmann$^1$,
J.~Jastrz\c{e}bski$^2$, 
B.~Ketzer$^1$, 
M.~Kisieli{\'n}ski$^2$,
B.~K{\l}os$^4$,
W.~Kurcewicz$^3$,
P.~Lubi{\'n}ski$^2$\footnote{Present addres: N. Copernicus Astronomical 
Center, Bartycka 18, 00-716 Warsaw, Poland},
P.~Napiorkowski$^2$,
L.~Pie{\'n}kowski$^2$,
R.~Smola{\'n}czuk$^5$,
E.~Widmann$^6$,
S.~Wycech$^5$}
\affiliation{$^1$Physik-Department, Technische Universit\"at M\"unchen,
                        D-85747 Garching, Germany}
\affiliation{$^2$Heavy Ion Laboratory, Warsaw University,
                                PL-02-093 Warsaw, Poland}

\affiliation{$^3$Institute of Experimental Physics, Warsaw University,
                                PL-00-681, Warsaw, Poland}

\affiliation{$^4$Physics Department, Silesian University,
                                PL-40-007 Katowice, Poland}
\affiliation{$^5$So{\l}tan Institute for Nuclear Studies,
                                PL-00-681 Warsaw, Poland}
\affiliation{$^6$CERN, CH-1211 Geneva 23, Switzerland}

\date{\today}

\begin{abstract}
The x-ray cascade from antiprotonic atoms was studied for $^{106}$Cd,
$^{116}$Cd, $^{112}$Sn, $^{116}$Sn, $^{120}$Sn, and $^{124}$Sn.
Widths and shifts of the levels due to strong interaction were
deduced.  Isotopic effects in the Cd and Sn isotopes are clearly
seen. The results are used to investigate the nucleon density in the
nuclear periphery.  The deduced neutron distributions are compared
with the results of the previously introduced radiochemical method and
with HFB calculations.
\end{abstract}


\maketitle

\section{INTRODUCTION}

Antiprotonic atoms are a specific tool to study the 
strong interaction and the nucleon density in the nuclear periphery. The
strong interaction potential leads to widths and energy shifts of
antiproton-atomic levels compared to the purely electromagnetic
interaction.  The measurement of these widths and shifts gives
information on the strength of the interaction, which is often
expressed by an effective scattering length in the optical potential
model~\cite{bat81b}, and on the nucleon density in the region where the
annihilation takes place. 

 In contrast to other methods which are
sensitive to the charge distribution and usually probe the whole
nucleus with the nuclear periphery giving only a small contribution,
antiprotons are sensitive to the matter density at the nuclear periphery 
(they probe the nucleon density at distances
about 2~fm larger than the half-denstity charge radius).
By investigating different isotopes of one element, the effect
of additional nucleons can be deduced. The major part of the effect
comes from the higher nucleon density in the nuclear periphery of
isotopes with more neutrons~\cite{koe86}. Possibly isospin effects on the
effective antiproton-nucleon scattering-length may also exist.

Before our studies data of antiprotonic atoms had been collected for
several elements~\cite{bat95}.  However, with a few exceptions, these
were mainly light isotopes ($Z < 40$) and a number of them was
investigated using natural targets under difficult antiprotonic beam
conditions. The aim of the PS209 collaboration was to measure with 
antiprotons from LEAR at CERN a large variety of elements and isotopes
in order to provide a set of data for a new combined
analysis~\cite{bat89} to determine the nucleon density in the nuclear
periphery. 
This analysis is expected to yield knowledge about the neutron density 
in the annihilation region and better
knowledge of the antiproton-nucleus interaction, e.g. about a density 
or isospin
dependence of the effective scattering length~\cite{bat95,wyc01}.

The results of the PS209 experiment were already reported in a number
of conferences, see e.g.~\cite{jas97,har01,trz01l}. In particular the
last reference presents a comprehensive table of level widths and
shifts determined in 34 monoisotopic or isotopically separated targets
ranging from \nuc{16}{O} to \nuc{238}{U}. In Ref.~\cite{trz01} these
data were analyzed under the assumption of a two parameter Fermi (2pF)
distribution of peripheral protons and neutrons. A linear relationship
of the difference between the neutron and proton root mean square
radii (rms) $\Delta r_{np}$ and the asymmetry parameter $\delta =
(N-Z)/A$ was established (where $N,Z$ and $A$ are neutron, atomic and
mass numbers, respectively). Besides conference communications some
more detailed reports on the evaluations of PS209 results were already
published~\cite{sch98,har02} or are in preparation.

In this publication results for tin and cadmium isotopes
are presented. The isotopes \nuc{106}{Cd}, \nuc{116}{Cd}, 
\nuc{112}{Sn}, \nuc{116}{Sn}, \nuc{120}{Sn} and \nuc{124}{Sn} have
been investigated. For four of these nuclei also 
the neutron-to-proton density ratio in the nuclear periphery could be measured
using the radiochemical method~\cite{jas93,lub94,lub98,sch99}.

\section{ EXPERIMENTAL METHOD AND SETUP }

The principle of the method employed is described in
Ref.~\onlinecite{sch98}.  The antiprotons are captured into a high
antiprotonic-atom orbit. They cascade down towards levels with lower
principal quantum number $n_{\overline{p}}$ by the emission of Auger
electrons and x-rays. In states with low $n_{\overline{p}}$ the orbit
of the antiproton comes close to the nucleus and the interaction with
the nucleus becomes large. The resulting shifts and widths of the
levels were partly evaluated and interpreted as explained in the next
sections.

The strong interaction width can be measured directly (via analysis of
the line shape) if it is of the order of magnitude of the instrumental
resolution (about 1~keV). For many isotopes this is the case for the
lowest visible transition. The energy of the transitions may be
measured with an accuracy of about 10~eV. Thus strong-interaction
energy shifts which are larger than this value may be determined. For
those levels, for which the strong interaction width is of the order
of the electromagnetic width (due to x- and Auger transitions), the
strong interaction width was deduced from the intensity balance of the
x-ray transitions feeding and depopulating the respective
level~\cite{koc68}. In the case of non-circular transitions the
feeding transitions cannot be observed experimentally, as they are
hidden by the much stronger intensities of the circular
transitions. In these cases the feeding intensities can be taken from
cascade calculations if the cascade is sufficiently well
known~\cite{sch98}.

\begin{figure}[hp]
\begin{center}
\includegraphics[width=0.7\textwidth]{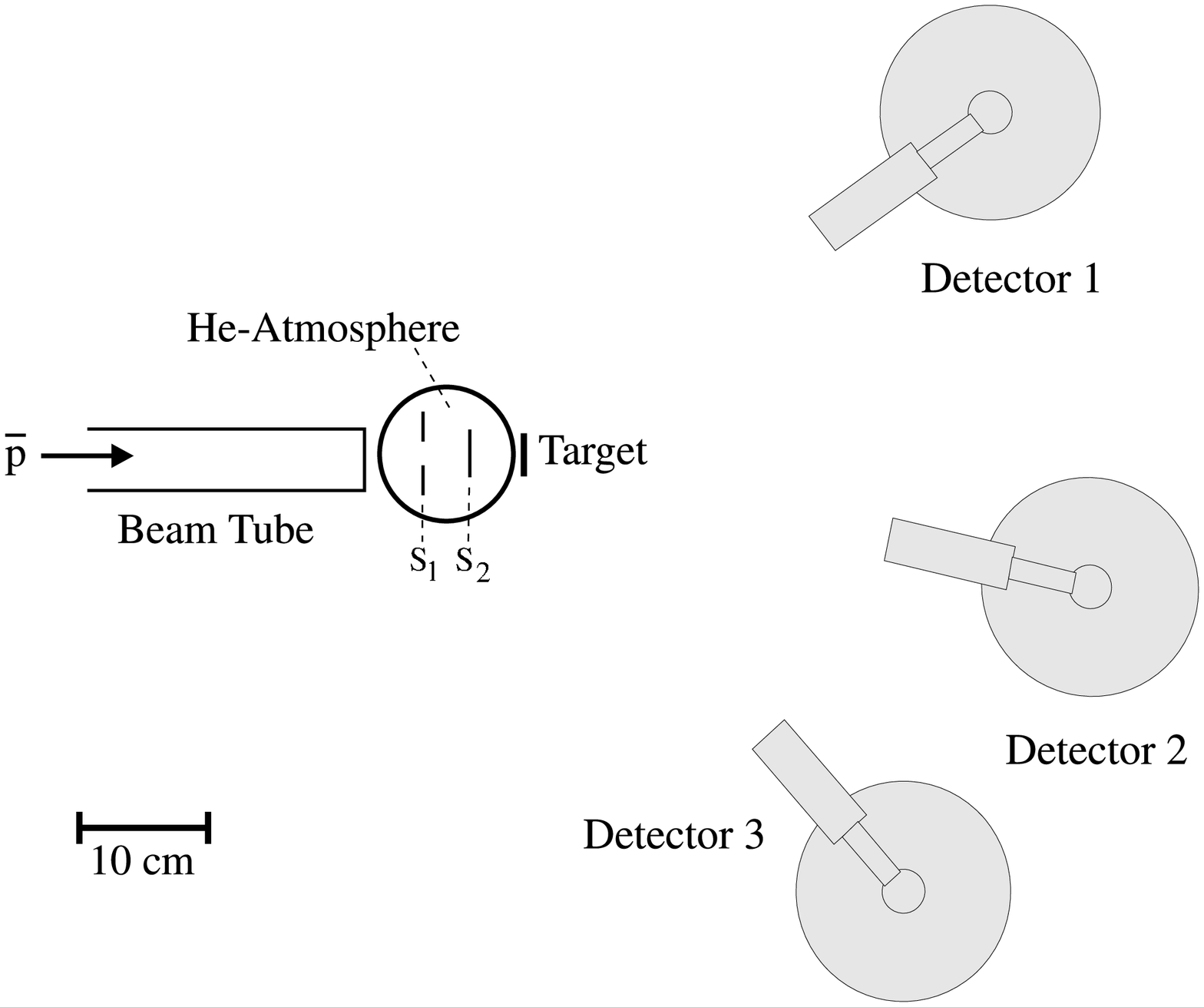} 
\end{center}
\caption{Schematic view of the experimental setup: S$_1$ -- anticounter, 
and S$_2$ -- counter of the telescope.}
\label{setup}
\end{figure}

The experiment was performed with the antiproton beam provided by LEAR
of CERN. The setup (cf. Fig.~\ref{setup}) is similar to that described
in Ref.~\onlinecite{sch98}. Due to the small initial momentum of the
antiprotons of 106~MeV/$c$ the scintillation-counter telescope,
consisting of an anticounter S1 and a counter S2, was placed inside a
chamber (with aluminum windows of thickness 12~$\mu$m) filled with
helium. This was necessary to avoid large an energy  
straggling and degradation for
the low energy (6.0~MeV) antiprotons in air. After passing the chamber
window the antiprotons were stopped inside the target. The properties of the
different targets are listed in Table~\ref{targets}.

The x-rays emitted during the antiproton cascade were measured with
three Ge detectors (two coaxial detectors with an active diameter of
49 mm and a length of 50~mm and one planar detector with diameter
36~mm and thickness 14~mm) with a resolution (FWHM) of about 1~keV at
200~keV gamma ray energy. The detectors were placed at distances of
about 50~cm from the target at angles of 13$^{\circ}$, 35$^{\circ}$
and 49$^{\circ}$ respectively towards the beam axis.  The
detector-target distance was adjusted in the way to obtain a good
signal-to-noise ratio and simultaneously decrease the background
produced by pions from the annihilation processes, which would obscure
the x-ray lines and would have damaged the detectors. This also
allowed to avoid summing effects. The x-rays were measured in
coincidence with the antiproton signal in a time window which was
extended up to 500~ns after the antiproton signal from the telescope
counter. The stability and efficiency of the detectors and the data
acquisition system was checked by on-line and off-line measurements
with calibration sources.

\section{ EXPERIMENTAL RESULTS}

\begin{figure}[hp]
\begin{center}
\includegraphics[width=8.6cm]{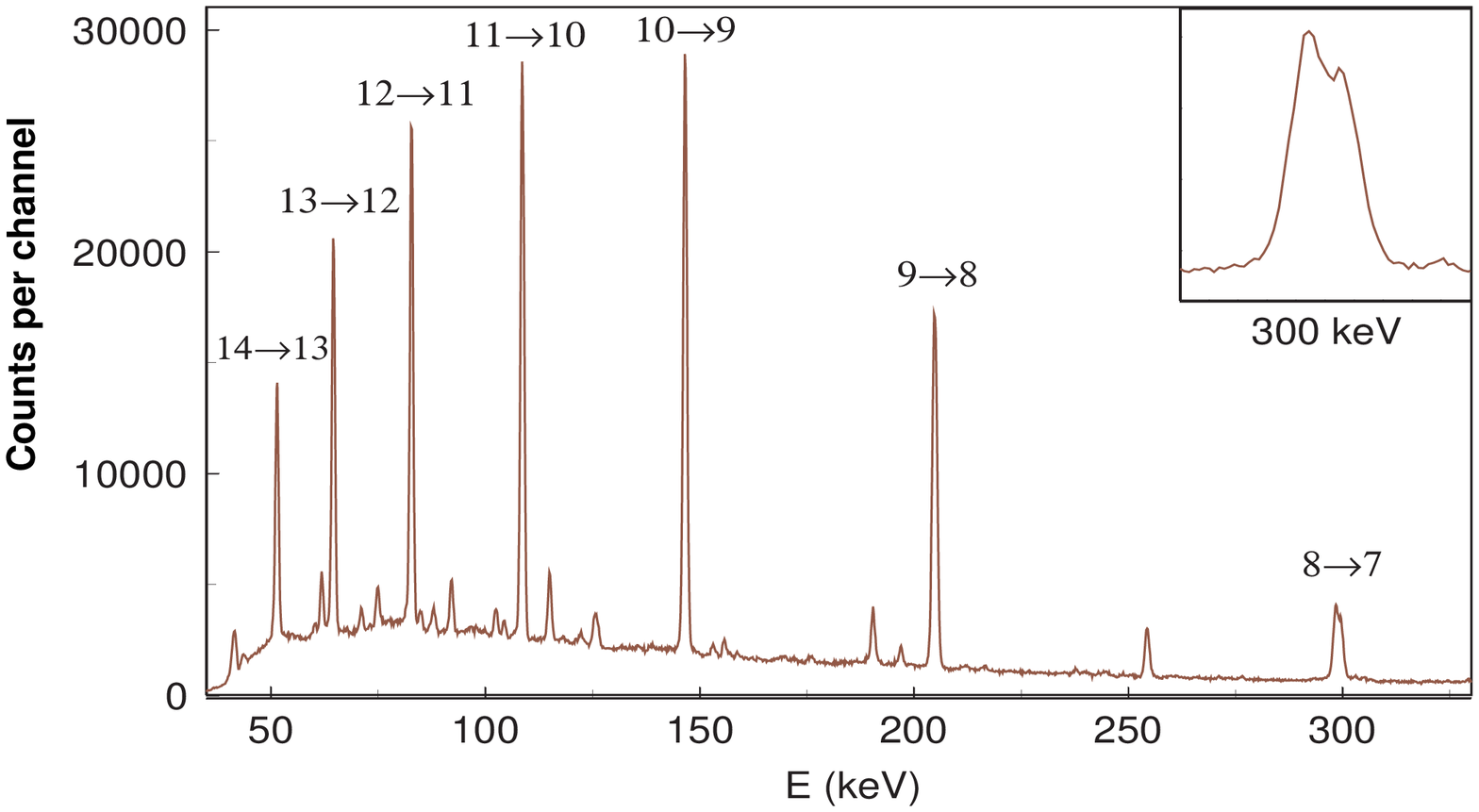} 
\end{center}
\caption{ Antiprotonic x-ray spectrum from $^{124}$Sn measured with
detector 1. The inset shows the spectrum around the the transition
$n=8\rightarrow7$.}
\label{sn124spec}
\end{figure}

The x-ray spectrum from the target \nuc{124}{Sn}, as taken with
detector~1, is shown in Fig.~\ref{sn124spec}. The upper right part shows the
part of the spectrum with the last visible transition
$n=8\rightarrow7$.

Those lines in the spectra which are not significantly broadened by
strong interaction were fitted with Gaussian functions. Their
relative intensities are given in Table~\ref{cd_int} and Table~\ref{sn_int}
for the Cd and Sn istopes, respectively. 
For the fit
of the transition $n=8\rightarrow7$ two Lorentzians convoluted with
Gaussians were used. 
The strong-interaction energy shifts were deduced
from the measured energies of these transitions.
The energy shift
is the difference between the energy 
calculated with a purely electromagnetic potential~\cite{bor83}
and the measured transition energy.

The widths of the levels $(n,l)=(8,7)$ were determined from the
measured intensity balance.  Small corrections for parallel
transitions and for unobserved transitions from higher levels were
taken from the calculated cascade~\cite{sch98}. For the determination
of the width of the level (9,7) all intensities of the feeding
transitions were taken from the results of the cascade
calculations. The radiative and Auger widths (obtainded according to
Ref.~\cite{leo74}) which were used for these calculations are
summarized in Table~\ref{cd_rad_aug} for cadmium and in
Table~\ref{sn_rad_aug} for tin.  Tables~\ref{cd_gam_sh}
and~\ref{sn_gam_sh} give the measured widths and shifts for the
cadmium and tin isotopes, respectively. The variation of these
observables due to the different number of protons and neutrons from
\nuc{106}{Cd} to \nuc{124}{Sn} is clearly visible. The widths for
\nuc{124}{Sn} are roughly twice as large as those for
\nuc{106}{Cd}. The shifts turn from attractive or compatible with zero
for \nuc{106}{Cd} to repulsive for \nuc{124}{Sn}.  The only observable
which does not follow rather smooth systematics is the upper
level $(n=8, l=7)$ width of \nuc{106}{Cd}. For all other nuclei
presented in these tables the lower to upper level widths ratio is
$\Gamma_{low}/\Gamma_{up} = 85 \pm 7$, whereas the same ratio is only
about 50 in case of \nuc{106}{Cd}.

This effect is due to E2 resonance~\cite{leo76} which in Cd nuclei
mixes the $n,l=6,5$ and the $n,l=8,7$ states.  The difference between
the energies of nuclear $2^+$ state and the corresponding
antiprotonic-atom transition is 65~keV and 184~keV in \nuc{106}{Cd}
and \nuc{116}{Cd}, respectively.  As the electric quadrupole moment is
not very different for both nuclei~\cite{ram87}, the increase of the
upper level width due to the mixing is more significant in
\nuc{106}{Cd} than in \nuc{116}{Cd}.  This qualitatively explains the
observed effect.

\begin{figure}[hp]
\begin{center}
\includegraphics[width=0.65\textwidth]{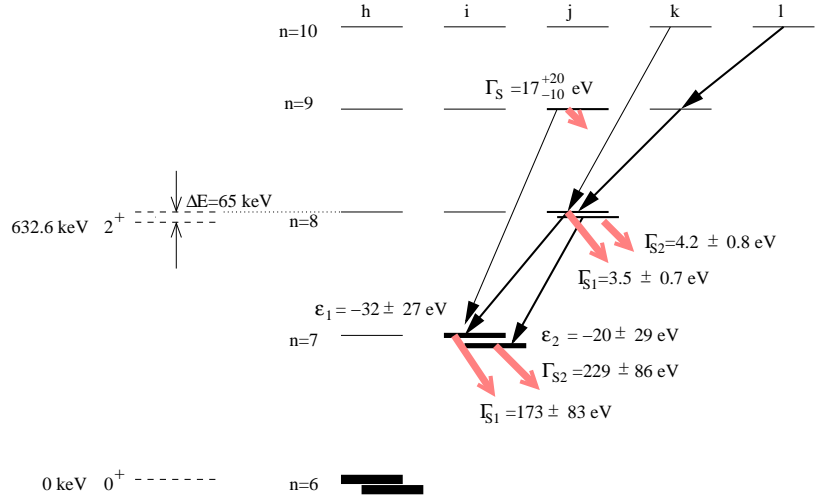}
\end{center}
\caption{Summary of measured shifts and widths for $^{106}$Cd and
the excitation energy of the nuclear 2$^+$ state in this nucleus.
(The upper level widths are the values before correction for the E2 effect).}
\label{cd_sum}
\end{figure}

To be more quantitative, the width of the $n,l=6,5$ level in Cd nuclei
should be known. This width was estimated by an extrapolation to
$Z=48$ of the systematics presented in~\cite{trz01zak} for lower Z
nuclei.  The extrapolated value is $7.7\pm 2.5$~keV.  This leads to
the E2 induced width of $1.9 \pm 0.5$~eV and $0.34\pm0.08$~eV in
\nuc{106}{Cd} and \nuc{116}{Cd}, respectively.  The $j^+$ and $j^-$
components of the upper level widths, corrected for the E2 effect, are
also given in Table~\ref{cd_gam_sh}.  The summary of the results
(measured values) for \nuc{106}{Cd} is shown in Fig.~\ref{cd_sum}

\section{ DISCUSSION }

The region of tin isotopes with a closed $Z = 50$ proton shell
constitutes one of the favorable parts of the Nuclear Chart for
experimental and theoretical nuclear-structure studies.  During our
investigation on antiprotonic atoms in this region we measured,
besides the results reported in this paper, the level widths and
shifts in even Te isotopes ($Z=52$)~\cite{klo02}.  In addition, using
the radiochemical method~\cite{jas93}, we have determined the neutron
halo factor, a quantity reflecting the composition of the outer
nuclear periphery in \nuc{106,116}{Cd},
\nuc{112,124}{Sn}~\cite{lub98,sch99} and in
\nuc{128,130}{Te}~\cite{lub98}.

In the present discussion we will concentrate on the first two
elements.  In our recent publication~\cite{trz01} we have presented in
detail our approach to determine the peripheral neutron distribution
and differences between the neutron and proton mean square radii
$\Delta r_{np}$ using observables gathered from antiprotonic atoms
under the assumption of a two-parameter Fermi (2pF) neutron and proton
distribution: $\rho(r) = \rho_0 \cdot \{1+exp(\frac{r-c}{a})\}^{-1}$,
where $c$ is the half density radius, $a$ the diffuseness parameter
(related to the surface thickness $t$ by $t=4\,\mbox{ln}3 \cdot a$)
and $\rho_0$ is a normalization factor.  This approach is summarized
below.

\begin{figure}[hp]
\begin{center}
\includegraphics[width=0.9\textwidth]{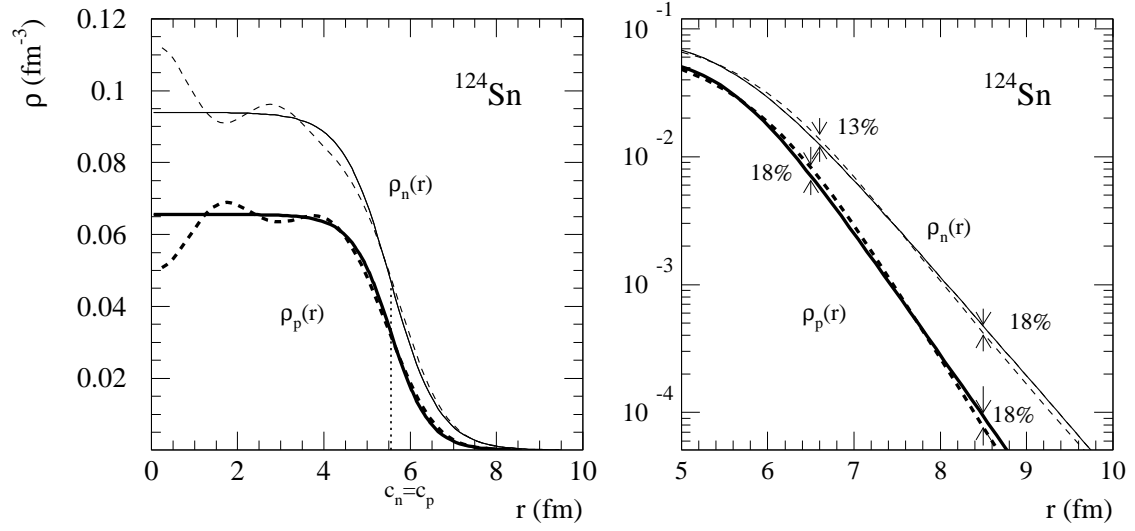} 
\end{center}
\caption{Comparison of the HFB model (dashed lines) and the two
parameter Fermi (2pF) density distributions (solid lines) for nucleus
\nuc{124}{Sn}. The 2pF distributions were fitted to HFB model curves
(half density radii $c_n = c_p = 5.55$~fm and the difference between
neutron and proton rms radii, $\Delta r_{np} = 0.16$~fm).  The
obtained 2pF diffuseness parameters are $a_p =0.45$~fm and
$a_n=0.57$~fm.}
\label{hfb_2pf}
\end{figure}

\begin{figure}[hp]
\begin{center}
\includegraphics[width=0.45\textwidth]{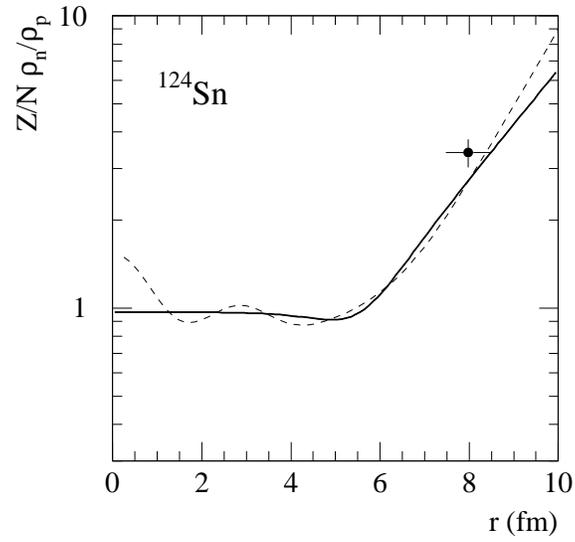} 
\end{center}
\caption{The same as for Fig.~\ref{hfb_2pf} but for the density ratio
($Z/N \cdot \rho_n / \rho_p$).  The cross indicates the halo factor
measured in the radiochemical experiment~\cite{sch99}.}
\label{hfb_2pf_rorat}
\end{figure}

Assuming identical annihilation probabilities on neutrons and
protons the radiochemical experiment determines the halo factor, which
is close to the normalized neutron to proton density ratio ($Z/N \cdot
\rho_n/ \rho_p$) at a radial distance $2.5 \pm 0.5$~fm larger than the
half density charge radius.  Comparing the halo factor with the
neutron to proton density ratio deduced from $\Delta r_{np}$
determined in other experiments one can conclude that for neutron rich
nuclei it is mostly the neutron diffuseness which increases and not
the half density radius~\cite{trz01}.  Although this conclusion was
based on the very simple 2pF model of the nuclear periphery it is
corroborated by much more sophisticated Hartree--Fock--Bogoliubov
(HFB) calculations. This is illustrated in Figs.~\ref{hfb_2pf}
and~\ref{hfb_2pf_rorat}, where the proton and neutron density
distributions for \nuc{124}{Sn} are compared for both models. The HFB
calculations were performed using SkP force~\cite{dob84}, giving
$\Delta r_{np}$ value equal to 0.16~fm. As the calculated proton
($c_p$) and neutron ($c_n$) half-density radii are almost identical,
this rms difference is mainly due to the difference in the proton and
neutron surface diffuseness.  The fitted 2pF distributions with the
HFB $c_n$, $c_p$ and $\Delta r_{np}$ values closely approximate the
HFB distributions. In the peripheral region from 6.5~fm to 8.5~fm,
e.g, the 2pF neutron distribution differs by less than 20\% from that
derived from HFB calculations. A similar result was obtained for other
investigated nuclei.

The antiprotonic x-rays are analyzed using an optical potential with
the antiproton-nucleon scattering length of the form $\overline{a}=
(2.5 \pm 0.3) \, + i (3.4 \pm 0.3)$~fm, as proposed for point-like
nucleons in Ref.~\cite{bat95}.  The method allows to study the nuclear
density at a radial distances about 1~fm closer to the nuclear center
than those examined in the radiochemical experiment.

The peripheral bare proton densities in form of 2pF distributions are
obtained~\cite{trz01} from the experiments sensitive to the nuclear
charge: electron scattering~\cite{vri87} or muonic
x-rays~\cite{fri95}.  The differences between experimental level
widths and shifts and those calculated with parameters of the proton
distributions are attributed to the neutron contributions to these
observables. Based on the analysis and the comparison described above,
the half density radii of the proton and neutron distributions are
assumed to be equal, $c_n=c_p$.  The neutron diffuseness is considered
as a free parameter, adjusted to agree best with the experimental
lower and upper level widths (the lower level shifts were not taken to
the fits, see comments below).

\begin{figure}[hp]
\begin{center}
\includegraphics[width=0.8\textwidth]{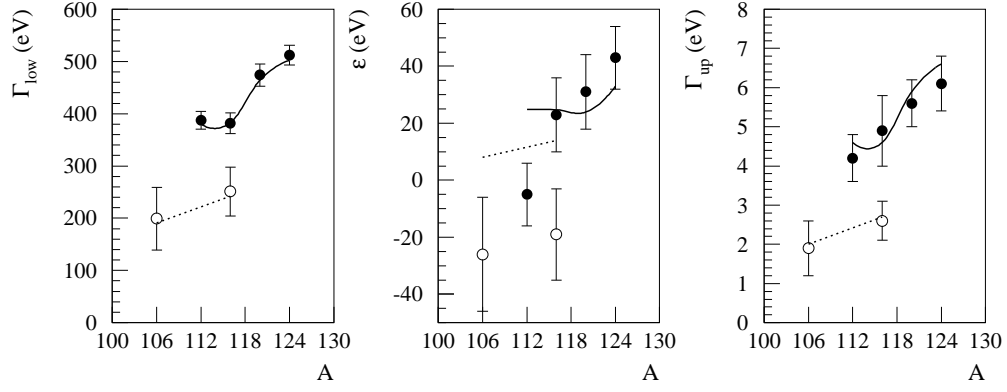}
\end{center}
\caption{Average widths and shifts of the levels (7,6)
and widths of the levels (8,7) plotted versus $A$.  Open circles and
dotted lines: cadmium isotopes (upper level widths corrected for the E2
effect), full circles and solid lines: tin isotopes. The lines are
calculated using the optical potential for point-like
nucleons~\cite{bat95} with the surface parameters given in
Table~\ref{dtpar} (see also text). Positive level shift corresponds to
repulsive interaction.}
\label{gam_sh_rys}
\end{figure}

Table~\ref{dtpar} illustrates this procedure for the Cd and Sn nuclei.
For the Sn nuclei the 2pF charge distribution determined using data
from muonic atoms or from electron scattering differ significantly.
Only the electron scattering data lead to $\Delta r_{np}$ values
compatible with the systematics gathered for other nuclei~\cite{trz01}
and with previous experiments~\cite{kra94,kra99}.  Therefore these
data were retained for further analysis.  In Fig.~\ref{gam_sh_rys} the
widths and shifts, calculated with density distributions from this
Table and the scattering lengths given above are compared with
corresponding experimental values. It is evident that the potential
used is able to reproduce simultaneously the lower and upper level
widths for Cd and Sn nuclei whereas one has some problems with the
level shifts (only for \nuc{116}{Sn}, \nuc{120}{Sn} and \nuc{124}{Sn}
the measured shifts are reproduced within the experimental errors).

\begin{figure}[hp]
\begin{center}
\includegraphics[width=0.8\textwidth]{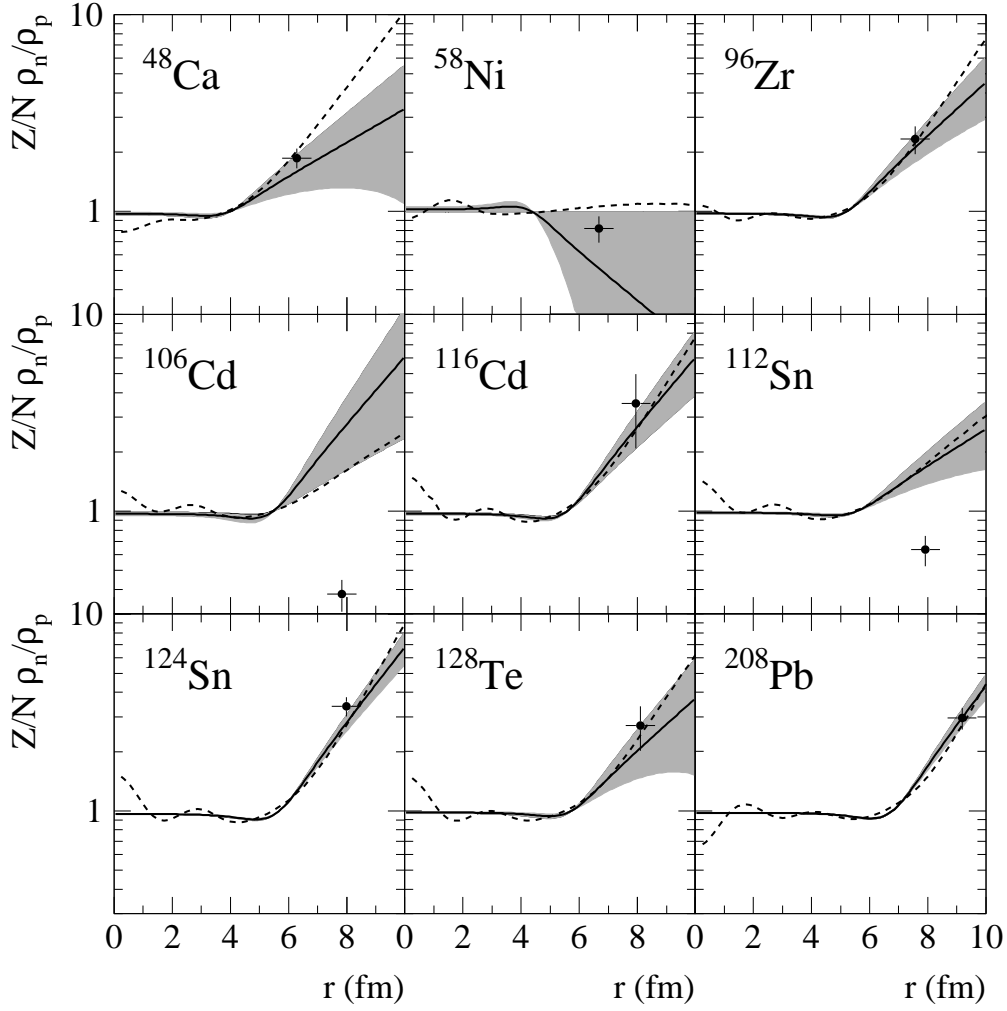}
\end{center}
\caption{Normalized neutron to proton density ratio 
($Z/N \cdot \rho_n / \rho_p$) deduced from strong-interaction level
widths and shifts (solid lines with indicated statistical errors) and
charge distributions given in Ref.~\cite{vri87} (Sn nuclei) and
Ref.~\cite{fri95} (other nuclei).  They are compared with $f_{halo}$
measured in the radiochemical experiments (marked with crosses at a
radial distance corresponding to the most probable annihilation site)
and with HFB model calculations (dashed lines).}
\label{rorat_prc}
\end{figure}

The analysis of the x-ray data as presented in Table~\ref{dtpar}
allows to determine the normalized neutron to proton density ratio
$Z/N \cdot \rho_n/\rho_p$ as a function of the radial distance at the
periphery of the investigated nuclei. As indicated above, the
radiochemical experiment can be considered as giving the same ratio at
a radial distance in the far periphery. Figure~\ref{rorat_prc}
compares the results of these two experiments together with the
normalized neutron to proton density ratio obtained from the
Hartree-Fock-Bogoliubov (HFB) calculations.  For the sake of
illustration the comparison is extended to some other nuclei not
discussed in detail in the present publication.
For heavy Cd and Sn nuclei two experimental approaches are consistent
within the experimental errors. They are also in fair agreement with
HFB calculations (a similar result is obtained for 15 other
investigated nuclei, partly shown in Fig.~\ref{rorat_prc}).

As already mentioned in our previous paper~\cite{trz01} the situation
is quite different for the lightest members of the Cd and Sn
chains. For these nuclei the analysis of the x-ray data gives
densities consistent with the HFB model with Skyrme interaction as
well as with recent calculations with Gogny force~\cite{pom00}. The
radiochemical experiment, however, seems to indicate a proton-rich
nuclear periphery. We encountered a similar problem for the two
lightest members of the Ru and Sm isotopic chains.  In
Ref.~\cite{wyc01} the role of a quasi-bound
$\overline{\mbox{p}}\mbox{p}$ ($^{13}\mbox{P}_0$) state in nuclei with
weakly bound protons was indicated as an explanation of this puzzle.
(For \nuc{106}{Cd} and \nuc{112}{Sn} the corresponding proton
separation energies are 7354~keV and 7559~keV, respectively).  The
formation of such a state would favor the annihilation on protons in
comparison with those on neutrons and lead to a much smaller halo
factor than really expected from the peripheral neutron and proton
densities. This explanation, although opening new research areas,
would indicate that our radiochemical method is not as universal as we
believed previously.

\begin{figure}[hp]
\begin{center}
\includegraphics[width=0.7\textwidth]{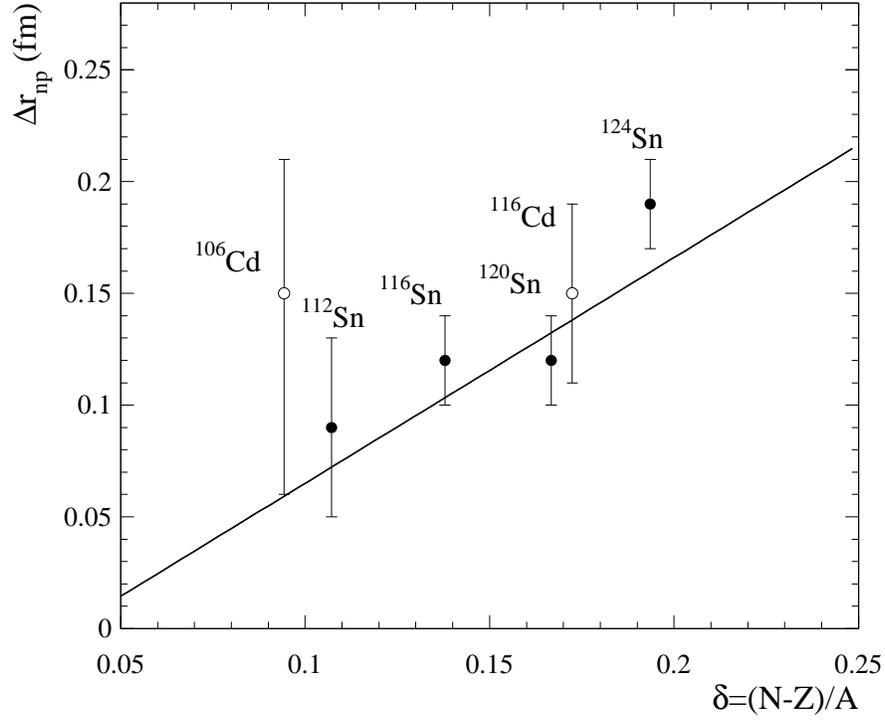}
\end{center}
\caption{Difference $\Delta r_{np}$ between the rms radii of the
neutron and proton distribution as deduced from the antiprotonic atom
x-ray data, as a function of $\delta = (N - Z) / A$. The full line is
the same as in Fig.~4 of Ref.~\cite{trz01}.}
\label{drdelta}
\end{figure}

The x-ray data, combined with proton distributions deduced from
electron scattering experiments (Sn nuclei) and muonic atoms (Cd
nuclei) allowed to determine the differences $\Delta r_{np}$ between
neutron and proton rms radii.  The results are presented in
Table~\ref{dtpar} and Fig.~\ref{drdelta}.  They are compared with
other experiments as well as with the HFB model in Ref.~\cite{trz01}.
The $\Delta r_{np}$ values in Figure~\ref{drdelta} for the
\nuc{116}{Cd} result differ from these in Ref.~\cite{trz01} as the
correction for E2 was done and the level shift was excluded from the
fit.  The \nuc{106}{Cd} results are presented for the first time.

\section{SUMMARY AND CONCLUSIONS}

Antiprotonic x-rays were measured in two even Cd and four even Sn
nuclei.  The strong interaction level widths and shifts were
determined. The observed isotopic effects are attributed, at least 
to a large extent, to the increase of the difference between the neutron
and proton rms radii with increasing neutron number.

The interpretation of the collected data was done using a simple two
parameter Fermi (2pF) model to describe the peripheral proton and
neutron distribution. It was verified that these simple distributions
approximate rather well (within 20\%) the distributions obtained from
the HFB model in the region where the antiproton annihilation
probability is significant. The parameters of the proton distributions
were obtained from literature, where 2pF charge distributions were
determined from muonic-atoms or electron-scattering experiments.

For neutron rich nuclei the peripheral neutron distributions deduced
from the antiprotonic x-ray data are in good agreement with ealier
radiochemical experiments. This is, however, not the case for the
lightest members of the investigated Cd and Sn isotope chains. In
these nuclei the radiochemical data indicate enhanced peripheral
proton density in comparison with the neutron density. Such a result
is in contradiction with the x-ray data as well as with HFB model
calculations.  It may be explained by the formation of a quasi-bound
\pbar p states in nuclei with weakly bound protons.

\begin{acknowledgments}
We thank the LEAR team for providing the
intense, high-quality antiproton beam and
Dr.~Anna~Stolarz of the Heavy Ion Laboratory in Warsaw and
Katharina~Nacke and
Dr.~Peter~Maier-Komor of the Technical University Munich
for the target preparation.
Financial support by 
the Polish State Committee
for Scientific Research as well as
the Accelerator Laboratory of the University and the
Technical University of Munich is acknowledged.
This work was also supported by Deutsche Forshungsgemeinschaft.
\end{acknowledgments}


\newpage 

\section*{TABLES}

\begin{table}[hb]
\caption{Target properties: thickness $d$, enrichment $a$,
          number of antiprotons used, 
         and on-line calibration sources.}
\begin{center}
\begin{ruledtabular}
\begin{tabular}{c|cccl}
Target & $d$ (mg/cm$^2$) & $a$ (\%)
                  & number of $\bar{p}$ ($10^8$) & calibration sources \\
\hline
$^{106}$Cd  & 40.0 & 76.5 & 9   & $^{137}$Cs, $^{152}$Eu \\
$^{116}$Cd  & 64.5 & 93.0 & 10  & $^{137}$Cs, $^{152}$Eu \\
$^{112}$Sn  & 65.6 & 94.7 & 17  & $^{137}$Cs, $^{152}$Eu \\
$^{116}$Sn  & 46.8 & 93.0 & 9   & $^{137}$Cs, $^{152}$Eu \\
$^{120}$Sn  & 65.3 & 99.2 & 11  & $^{137}$Cs, $^{152}$Eu \\
$^{124}$Sn  & 70.1 & 97.9 & 23  & $^{133}$Ba, $^{137}$Cs, $^{152}$Eu \\
\end{tabular}
\end{ruledtabular}
\end{center}
\label{targets}
\end{table}

\begin{table}
\caption{Measured relative antiprotonic x-ray intensities normalized to the
	 transition $n=11\rightarrow10$ (mean values of the results from 
	 three detectors).}
\resizebox{!}{10cm}{
\begin{ruledtabular}
\begin{tabular}{cc|r|cc}
Transitions        &        &        Energy &  $^{106}$Cd            
                                            &  $^{116}$Cd \\
		     &        &      [keV]  &                        
                                            &             \\   
\hline
 8$\rightarrow$7   &                  & 276 
&     72.70  $\pm$  2.79 &   75.64  $\pm$ 2.84 \\
 9$\rightarrow$8   &                  & 188 
&     119.01 $\pm$  6.24 &   114.53 $\pm$ 5.80 \\
10$\rightarrow$9   &13$\rightarrow$11 & 135 
&     131.46 $\pm$  6.62 &   132.17 $\pm$ 6.98 \\
11$\rightarrow$10  &                  & 100 
&     100.00 $\pm$  5.04 &   100.00 $\pm$ 5.81 \\
12$\rightarrow$11  &                  &  76 
&      83.28 $\pm$  4.21 &   84.46  $\pm$ 6.95 \\
13$\rightarrow$12  &                  &  59 
&      66.35 $\pm$  3.42 &   66.66  $\pm$ 8.10 \\
14$\rightarrow$13  &18$\rightarrow$16 &  47 
&      54.54 $\pm$  2.99 &   56.03  $\pm$ 10.9 \\
		   &                  &     
&                        &                     \\
 9$\rightarrow$7   &                  & 464 
&       5.38 $\pm$  0.95 &   5.00   $\pm$ 0.64 \\
10$\rightarrow$8   &                  & 323 
&      11.71 $\pm$  0.74 &   11.53  $\pm$ 0.76 \\
11$\rightarrow$9   &13$\rightarrow$10 & 234 
&      22.78 $\pm$  1.20 &   20.97  $\pm$ 1.12 \\
12$\rightarrow$10  &                  & 175 
&      18.40 $\pm$  3.61 &   17.30  $\pm$ 0.92 \\
14$\rightarrow$12  &                  & 106 
&      13.64 $\pm$  0.74 &   13.96  $\pm$ 0.86 \\
15$\rightarrow$13  &                  &  84 
&      10.27 $\pm$  0.58 &   10.54  $\pm$ 0.77 \\
16$\rightarrow$14  &                  &  68 
&       6.10 $\pm$  0.38 &   7.18   $\pm$ 0.74 \\
17$\rightarrow$15  &                  &  56 
&      12.09 $\pm$  0.68 &   10.99  $\pm$ 1.52 \\
19$\rightarrow$17  &                  &  39 
&       9.97 $\pm$  1.0  &   18.38  $\pm$ 7.20 \\
		     &                 &     
&                        &                     \\
11$\rightarrow$8   & 7$\rightarrow$6  & 423 
&       5.59 $\pm$  0.69 &   3.72   $\pm$ 0.53 \\
12$\rightarrow$9   &                  & 310 
&       3.81 $\pm$  0.39 &   4.41   $\pm$ 0.43 \\
14$\rightarrow$11  &                  & 181 
&       5.09 $\pm$  0.92 &   5.72   $\pm$ 0.37 \\
15$\rightarrow$12  &                  & 143 
&       4.22 $\pm$  0.31 &   4.2    $\pm$ 0.5  \\
16$\rightarrow$13  &18$\rightarrow$14 & 115 
&       5.16 $\pm$  0.36 &   5.20   $\pm$ 0.37 \\
17$\rightarrow$14  &                  &  94 
&       6.29 $\pm$  0.40 &   6.23   $\pm$ 0.48 \\
18$\rightarrow$15  &                  &  78 
&       3.84 $\pm$  0.33 &   4.83   $\pm$ 0.45 \\
19$\rightarrow$16  &                  &  65 
&       2.5  $\pm$  0.5  &   2.82   $\pm$ 0.37 \\
		     &                &     
&                        &                     \\
12$\rightarrow$8   &                  & 498 
&       1.28 $\pm$  0.46 &   1.30   $\pm$ 0.53 \\
13$\rightarrow$9   &                  & 369 
&       1.30 $\pm$  0.38 &   2.17   $\pm$ 0.33 \\
14$\rightarrow$10  &                  & 281 
&       1.86 $\pm$  0.55 &   1.83   $\pm$ 0.25 \\
15$\rightarrow$11  &                  & 219 
&       2.33 $\pm$  0.31 &   1.83   $\pm$ 0.25 \\
16$\rightarrow$12  &                  & 174 
&       0.99 $\pm$  0.79 &   1.81   $\pm$ 0.36 \\
17$\rightarrow$13  &                  & 141 
&       2.6  $\pm$  0.26 &   2.81   $\pm$ 0.26 \\
19$\rightarrow$15  &                  &  96 
&       2.73 $\pm$  0.26 &   2.7    $\pm$ 0.5  \\
		     &                &     
&                        &                     \\
17$\rightarrow$12  &                  & 200 
&       2.0  $\pm$  0.5  &   2.01   $\pm$ 0.27 \\
18$\rightarrow$13  &                  & 162 
&       1.79  $\pm$ 0.24 &   1.69   $\pm$ 0.23 \\
19$\rightarrow$14  &                  & 133 
&       1.97  $\pm$ 0.32 &   1.81   $\pm$ 0.75 \\
\end{tabular}
\end{ruledtabular}
}
\label{cd_int}
\end{table}

\begin{table}[hb]
\caption{Measured relative antiprotonic x-ray intensities normalized to the
	   transition $n=11\rightarrow10$ (mean values of the results from 
	   three detectors).}
\resizebox{!}{10cm}{
\begin{ruledtabular}
\begin{tabular}{cc|r|cccc}
 Transitions       &            &  Energy &       $^{112}$Sn  & $^{116}$Sn  
                                          &       $^{120}$Sn  & $^{124}$Sn \\
		   &            &   [keV] &                   &              
                   &                    & \\
\hline
 8$\rightarrow$7  &                & 299& 70.71 $\pm$ 2.67& 65.35 $\pm$ 4.90 
                                        & 60.82 $\pm$ 2.20& 56.19 $\pm$ 2.51 \\
 9$\rightarrow$8  &                & 205&114.64 $\pm$ 5.77& 114.72 $\pm$ 5.78 
                                        &113.10 $\pm$ 5.76& 110.00 $\pm$ 5.53 \\
10$\rightarrow$9 &13$\rightarrow$11& 146&128.09 $\pm$ 6.58& 125.66 $\pm$ 6.38 
                                        &126.99 $\pm$ 6.51& 126.26 $\pm$ 6.48 \\
11$\rightarrow$10 &                & 108&100.00 $\pm$ 5.44& 100.00 $\pm$ 5.10 
                                        &100.00 $\pm$ 5.43& 100.00 $\pm$ 5.61 \\
12$\rightarrow$11 &                & 82 & 82.68 $\pm$ 5.26&  83.21 $\pm$ 4.50 
                                        & 84.06 $\pm$ 5.33&  83.95 $\pm$ 5.75 \\
13$\rightarrow$12 &                & 64 & 68.96 $\pm$ 6.27&  70.42 $\pm$ 4.45 
                                        & 72.49 $\pm$ 6.61&  72.97 $\pm$ 7.26 \\
13$\rightarrow$14 &18$\rightarrow$16&51 & 57.39 $\pm$ 7.57&  59.83 $\pm$ 4.93 
                                        & 61.39 $\pm$ 8.05&  61.55 $\pm$ 8.85 \\
15$\rightarrow$14 &                 & 41& 26.38 $\pm$ 5.34&  27.33 $\pm$ 3.26 
                                        & 31.55 $\pm$ 6.33&  29.47 $\pm$ 6.52 \\
                  &                 &   &                 &                    
                                        &                 &                   \\
 9$\rightarrow$7  &                 &503&  4.15 $\pm$ 0.31&   4.56 $\pm$ 1.0  
                                        &  3.69 $\pm$ 0.34&   3.52 $\pm$ 0.26 \\
10$\rightarrow$8  &                 &350& 12.03 $\pm$ 0.67&  11.55 $\pm$ 0.64 
                                        & 11.81 $\pm$ 0.69&  11.56 $\pm$ 0.77 \\
11$\rightarrow$9  &13$\rightarrow$10&255& 18.92 $\pm$ 1.17&  19.54 $\pm$ 1.99 
                                        & 18.83 $\pm$ 2.05&  16.86 $\pm$ 1.50 \\
12$\rightarrow$10 &16$\rightarrow$12&190& 14.41 $\pm$ 0.75&  13.96 $\pm$ 0.73 
                                        & 14.15 $\pm$ 0.74&  13.42 $\pm$ 0.70 \\
14$\rightarrow$12 &                 &115& 12.00 $\pm$ 0.67&  12.02 $\pm$ 0.64 
                                        & 12.26 $\pm$ 0.68&  12.40 $\pm$ 6.92 \\
15$\rightarrow$13 &                 & 92&  8.54 $\pm$ 0.54&   8.72 $\pm$ 0.48 
                                        &  8.74 $\pm$ 0.54&   8.49 $\pm$ 0.54 \\
16$\rightarrow$14 &                 & 74&  6.57 $\pm$ 0.53&   6.56 $\pm$ 0.41 
                                        &  5.29 $\pm$ 0.52&   6.20 $\pm$ 0.59 \\
17$\rightarrow$15 &                 & 61& 10.49 $\pm$ 1.04&  11.23 $\pm$ 0.78 
                                        & 11.56 $\pm$ 1.12&  11.79 $\pm$ 1.29 \\
19$\rightarrow$17 &                 & 43&  6.59 $\pm$ 1.25&   5.73 $\pm$ 0.71 
                                        &  8.38 $\pm$ 1.65&   6.26 $\pm$ 1.32 \\
		  &                 &   &                 &                    
                                        &                 &                   \\
11$\rightarrow$8  &                 &458&  2.03 $\pm$ 0.20&   1.7  $\pm$ 0.5
                                        &  1.76 $\pm$ 0.19&   1.58 $\pm$ 1.0  \\
12$\rightarrow$9  &                 &336&  4.13 $\pm$ 0.29&   4.11 $\pm$ 0.31 
                                        &  3.86 $\pm$ 0.27&   3.16 $\pm$ 0.22 \\
14$\rightarrow$11 &                 &197&  4.53 $\pm$ 0.29&   3.89 $\pm$ 0.32 
                                        &  4.43 $\pm$ 0.28&   4.68 $\pm$ 0.26 \\
15$\rightarrow$12 &                 &156&  3.5  $\pm$ 0.5 &   3.0  $\pm$ 0.5  
                                        &  3.75 $\pm$ 0.22&   3.62 $\pm$ 0.21 \\
16$\rightarrow$13 &18$\rightarrow$14&125&  3.71 $\pm$ 0.27&   3.92 $\pm$ 0.25 
                                        &  4.15 $\pm$ 0.25&   3.9  $\pm$ 1.0  \\
17$\rightarrow$14 &                 &102&  5.05 $\pm$ 0.31&   4.60 $\pm$ 0.27 
                                        &  4.47 $\pm$ 0.27&   4.83 $\pm$ 0.30 \\
18$\rightarrow$15 &                 & 84&  3.20 $\pm$ 0.24&   3.18 $\pm$ 0.22 
                                        &  3.29 $\pm$ 0.24&   2.76 $\pm$ 0.22 \\
19$\rightarrow$16 &                 & 71&  2.49 $\pm$ 0.23&   2.86 $\pm$ 0.19 
                                        &  3.14 $\pm$ 0.28&   3.61 $\pm$ 0.33 \\
                  &                 &   &                 &                 
                                        &                 &                   \\
13$\rightarrow$9  &                 &400&  1.56 $\pm$ 0.18&   1.72 $\pm$ 0.19 
                                        &  1.61 $\pm$ 0.16&   1.15 $\pm$ 0.13 \\
14$\rightarrow$10 &                 &305&  1.12 $\pm$ 0.14&   0.98 $\pm$ 0.16 
                                        &  1.41 $\pm$ 0.15&   1.18 $\pm$ 0.17 \\
15$\rightarrow$11 &                 &238&  1.92 $\pm$ 0.15&   1.83 $\pm$ 0.17 
                                        &  1.49 $\pm$ 0.14&   1.69 $\pm$ 0.14 \\
17$\rightarrow$13 &                 &153&  2.43 $\pm$ 0.17&   2.13 $\pm$ 0.16 
                                        &  1.96 $\pm$ 0.14&   2.24 $\pm$ 0.15 \\
19$\rightarrow$15 &                 &104&  1.45 $\pm$ 0.14&   1.33 $\pm$ 0.13 
                                        &  1.28 $\pm$ 0.12&   1.3  $\pm$ 1.0  \\
\end{tabular}
\end{ruledtabular}
}
\label{sn_int}
\end{table}

\begin{table}[hb]
\caption{ Radiative width $\Gamma_{\rm em}$ and Auger width
            $\Gamma_{\rm Auger}$ for those levels of $\bar{\rm p}$Cd
            where the strong interaction width was determined via the
            intensity balance. Values in eV.}
\begin{ruledtabular}
\begin{tabular}{c|cc|cc}
       & \multicolumn{2}{c|}{$^{106}$Cd} & \multicolumn{2}{c}{$^{116}$Cd} \\
\raisebox{1.5ex}[-1.5ex]{$(n,l)$} & $\Gamma_{\rm em}$ &
$\Gamma_{\rm Auger}$  & $\Gamma_{\rm em}$  & $\Gamma_{\rm Auger}$ \\ 
\hline
(8,7) & 4.95 & 0.04 & 4.70 & 0.04 \\
(9,7) & 3.49 & 0.06 & 3.31 & 0.05 \\
(7,6) & 9.86 & 0.03 &      &      \\
\end{tabular}
\end{ruledtabular}
\label{cd_rad_aug}
\end{table}

\begin{table}[hb]
\caption{Radiative width $\Gamma_{\rm em}$ and Auger width
            $\Gamma_{\rm Auger}$ for those levels of $\bar{\rm p}$Sn
            where the strong interaction width was determined via the
            intensity balance. Values in eV.} 
\begin{ruledtabular}
\begin{tabular}[t]{c|cc|cc|cc|cc}
       & \multicolumn{2}{c|}{$^{112}$Sn} & \multicolumn{2}{c|}{$^{116}$Sn} &
         \multicolumn{2}{c|}{$^{120}$Sn} & \multicolumn{2}{c}{$^{124}$Sn} \\
\raisebox{1.5ex}[-1.5ex]{$(n,l)$} & $\Gamma_{\rm em}$ &    
$\Gamma_{\rm Auger}$ & $\Gamma_{\rm em}$ & $\Gamma_{\rm Auger}$ &
$\Gamma_{\rm em}$ & $\Gamma_{\rm Auger}$ & $\Gamma_{\rm em}$ &
$\Gamma_{\rm Auger}$ \\
\hline
(8,7) & 5.79 & 0.04 & 5.67 & 0.04 & 5.56 & 0.04 & 5.46 & 0.04 \\
(9,7) & 4.08 & 0.06 & 3.99 & 0.06 & 3.92 & 0.06 & 3.85 & 0.05 \\
\end{tabular}
\end{ruledtabular}
\label{sn_rad_aug}
\end{table}

\begin{table}[hb]
\caption{Measured level widths and shifts for the cadmium isotopes.}
\label{cd_gam_sh}
\begin{ruledtabular}
\begin{tabular}{c|cc|cc}
       & \multicolumn{2}{c|}{$^{106}$Cd} & \multicolumn{2}{c}{$^{116}$Cd} \\
       & $j=l+1/2$ &
$j=l-1/2$ & $j=l+1/2$ & $j=l-1/2$ \\ 
\hline
$\Gamma(7,6)$ (eV)     & 173$\pm$83  & 229$\pm$86 & 307$\pm$63 & 186$\pm$69 \\
$\varepsilon(7,6)$ (eV)& -32$\pm$27  & -20$\pm$29 & -15$\pm$22 & $-24\pm24$ \\
$\Gamma(8,7)$ (eV)     & 3.5$\pm$0.7 & 4.2$\pm$0.8& 2.7$\pm$0.6& 3.3$\pm$0.7 \\
$\Gamma(8,7)^*$ (eV)   & 1.4$\pm$1.0 & 2.4$\pm$1.0& 2.4$\pm$0.6& 3.0$\pm0.7$ \\
$\Gamma(9,7)$ (eV)     & \multicolumn{2}{c|}{17$^{+20}_{-10}$}&
                         \multicolumn{2}{c}{18$^{+19}_{-7}$} \\ \hline
\multicolumn{5}{l}{$^*$ After the correction for the E2 effect (see text).}
\end{tabular}
\end{ruledtabular}
\end{table}

\begin{table}[hb]
\caption{ Measured level widths and shifts for the tin isotopes.}
\label{sn_gam_sh}
{\scriptsize
\begin{ruledtabular}
\begin{tabular}{c|cc|cc|cc|cc}
       & \multicolumn{2}{c|}{$^{112}$Sn} & \multicolumn{2}{c|}{$^{116}$Sn} &
         \multicolumn{2}{c|}{$^{120}$Sn} & \multicolumn{2}{c}{$^{124}$Sn} \\
       & $j=l+1/2$ &
$j=l-1/2$ & $j=l+1/2$ & $j=l-1/2$ &
$j=l+1/2$ & $j=l-1/2$ & $j=l+1/2$ &
$j=l-1/2$ \\ 
\hline
$\Gamma(7,6)$ (eV)     & 411$\pm$22 & 358$\pm$25 & 386$\pm$27 & 377$\pm$31 &
                         448$\pm$27 & 505$\pm$32 & 493$\pm$25 & 534$\pm$29 \\
$\varepsilon(7,6)$ (eV)& -9$\pm$16   & -1$\pm$13   & $12\pm18$ & $36\pm19$ &
                         $26\pm17$ & $37\pm20$ & $26\pm17$ & $63\pm16$ \\
$\Gamma(8,7)$ (eV)     & 4.1$^{+0.8}_{-0.7}$ & 4.3$^{+0.8}_{-0.7}$ &
                         4.7$^{+1.4}_{-1.1}$ & 5.2$^{+1.5}_{-1.2}$ &
                         4.9$^{+0.8}_{-0.7}$ & 6.4$^{+1.0}_{-0.8}$ &
                         5.5$^{+1.0}_{-0.9}$ & 6.8$^{+1.1}_{-1.0}$ \\
$\Gamma(9,7)$ (eV)     & \multicolumn{2}{c|}{20$^{+13}_{-6}$} &
                         \multicolumn{2}{c|}{17$^{+12}_{-6}$} &
                         \multicolumn{2}{c|}{22$^{+12}_{-6}$} &
                         \multicolumn{2}{c}{24$^{+15}_{-7}$} \\
\end{tabular}
\end{ruledtabular}
}
\end{table}

\begin{turnpage}
\begin{table}[hb]
\caption{Parameters of 2pF neutron density distributions deduced from
the widths of antiprotonic levels in Cd and Sn atoms.}
\begin{ruledtabular}
\begin{tabular}{c|l|l|l|l|r|r|r|l|l|l|l|r|r|r} 
  &\multicolumn{7}{c|}{Charge distributions from muonic atoms$^{a)}$}
  &\multicolumn{7}{c}{ Charge distributions from electron scattering$^{b)}$} \\
   \hline 
  Isotope  &$c_{ch}$&$t_{ch}$&$c_{p}$&$t_{p}$&$\Delta t_{np}$&$\chi^2$
&$\Delta r_{np}$&
            $c_{ch}$&$t_{ch}$&$c_{p}$&$t_{p}$&$\Delta t_{np}$&$\chi^2$&
$\Delta r_{np}$\\ \hline
$^{106}$Cd & 5.2875 &  2.30  & 5.329 & 2.000 &$^{c)}$ 0.43$^{+0.18}_{-0.25}$
& 0.2 & 0.15$^{+0.06}_{-0.09}$
           &        &        &       &       &                       
&           &     \\
$^{116}$Cd & 5.4164 &  2.30  & 5.457 & 2.000 &$^{c)}$ 0.45$^{+0.10}_{-0.13}$
& 0.2 & 0.15$\pm$0.4
           &   5.42 &  2.34  & 5.461 & 2.043 & 0.39$^{+0.11}_{-0.13}$& 0.2 
& 0.13$^{+0.05}_{-0.09}$ \\
$^{112}$Sn & 5.3714 &  2.30  & 5.412 & 1.995 & 0.53$\pm$0.06         & 0.8 
& 0.19$\pm$0.02  
           & 5.375  &  2.416 & 5.416 & 2.184 & 0.26$\pm$0.07         & 0.8 

& 0.09$\pm$0.04 \\
$^{116}$Sn & 5.417  &  2.30  & 5.458 & 1.995 & 0.48$\pm$0.07         & 0.3
       & 0.17$\pm$0.02 
           & 5.358  &  2.420 & 5.399 & 2.135 & 0.33$\pm$0.06         & 0.3 
      & 0.12$\pm$0.02  \\  
$^{120}$Sn & 5.459  &  2.30  & 5.499 & 1.995 & 0.58$\pm$0.05         & 0.7
       & 0.20$\pm$0.02
           & 5.315  &  2.530 & 5.356 & 2.263 & 0.33$\pm$0.06         & 0.9
       & 0.12$\pm$0.02 \\
$^{124}$Sn & 5.491  &  2.30  & 5.531 & 1.995 & 0.60$\pm$0.06         & 0.9
       & 0.21$\pm$0.02
           & 5.490  &  2.347 & 5.530 & 2.052 & 0.53$\pm$0.06         & 0.9
       & 0.19$\pm$0.02 \\
\multicolumn{15}{l}{$^{a)}$ Reference~\cite{fri95}.} \\
\multicolumn{15}{l}{$^{b)}$ Reference~\cite{vri87}.} \\
\multicolumn{15}{l}{$^{c)}$ the upper widths corrected for E2 effect 
were used in the fit.} \\
\multicolumn{15}{l}{$c_{ch},t_{ch}$ -- the half-density radius and 
the surface thickness of 
charge density distributions.}\\
\multicolumn{15}{l}{ $c_{p},t_{p}$ -- the half-density radius and 
the surface thickness of point-like proton density distributions.}\\ 
\multicolumn{15}{l}{$\Delta t_{np}$ -- difference of the surface 
thicknesses of proton and neutron distributions.} \\
\end{tabular}
\end{ruledtabular}
\label{dtpar}
\end{table}
\end{turnpage}

\clearpage

\section*{FIGURES}

\clearpage

\clearpage

\clearpage

\clearpage

\clearpage

\clearpage

\end{document}